\documentclass[aps,showpacs,groupedaddress]{revtex4}  
\usepackage{graphicx}  
\usepackage{dcolumn}   
\usepackage{bm}        
\usepackage{amssymb}   
\usepackage{amsmath}
\usepackage{epsfig}
\usepackage{xspace}
\usepackage{color}
\usepackage[utf8]{inputenc}

\newcommand{\bea}{\begin{eqnarray}}
\newcommand{\eea}{\end{eqnarray}}

\newcommand{\Tr}{\textrm{Tr}}

\begin{document}


\title{Calculating the EoS of the dense quark-gluon plasma using the Complex Langevin equation}

\author{D\'enes Sexty }

\affiliation{ {\it Department of Physics, Wuppertal University, Gaussstr. 20, D-42119 Wuppertal, Germany}; {\it J\"ulich Supercomputing Centre, Forschungszentrum J\"ulich, D-52425 J\"ulich, Germany} }


\begin{abstract} 
  The pressure and energy density of the quark-gluon plasma at
  finite baryon chemical
  potential are calculated using the Complex Langevin equation. 
  The stout smearing procedure is generalized for the SL(3,$\mathcal{C}$)
  manifold allowing the usage of an improved action in the Complex
  Langevin setup. Four degenerate flavors of staggered quarks
  with $m_\pi=500-700$ MeV
  are used with a tree-level Symanzik improved gauge
  action on $16^3 \times 8 $ lattices.
    Results are compared to the Taylor expansion and good agreement is found for small chemical potentials. 
\end{abstract}
\pacs{11.15.Ha, 12.38.Gc}

\maketitle

\section{Introduction and overview}\label{sec:introduction}

The strong interactions are described by Quantum Chromodynamics (QCD).
The determination of the QCD phase diagram is one of the great challenges of
the theoretical study of this theory.
It is phenomenologically relevant in many areas, such as in
the early Universe, in relativistic heavy-ion collision experiments as well
as in astronomy describing neutron stars.
The lattice discretisation of the theory allows for the precise calculation 
of the Equation of State (EoS),  the exploration of the
hadronic and the quark-gluon plasma phase at zero baryonic density
and the phase transition between them \cite{Petreczky:2012rq,Philipsen:2012nu,Borsanyi:2016bzg}.
The EoS of the QCD matter can be calculated on the lattice
using various methods. Usually the partition function is constructed by
measuring its derivatives and integrating from a starting point
at vacuum \cite{Engels:1990vr,Boyd:1996bx}, but other approaches are also
available \cite{Meyer:2009tq,Giusti:2010bb,Suzuki:2013gza,Caselle:2016wsw}.
However, the lattice formulation of QCD suffers from a problem
at nonzero chemical potential:
the partition sum of the theory 
is written in terms of a complex measure 
due to the fermion determinant,
thus the standard importance sampling approaches are invalid.
This is called the QCD sign problem. 
Several different strategies have been proposed 
to circumvent the sign problem (see reviews 
\cite{Fodor:2009ax,deForcrand:2010ys,Aarts:2013bla,Sexty:2014dxa}).
The EoS at $ \mu>0$ is traditionally calculated via Taylor expansion
or reweighting \cite{Fodor:2002km,Allton:2002zi,Allton:2003vx,Allton:2005gk,Borsanyi:2012cr,Bazavov:2017dus}
and analytical continuation \cite{DElia:2009pdy,Gunther:2016vcp}.
In this study the Complex Langevin equation (CLE) is used to carry
out simulations directly at $\mu>0$.

The complexification of the Langevin equation 
was proposed long ago \cite{Parisi:1984cs,Klauder:1983nn}. 
The idea is to circumvent the sign problem by complexifing the field
manifold of the theory and defining a stochastic process on this
manifold using analiticity.
After an initial excitement it was noticed that the 
Complex Langevin equation sometimes gives wrong results \cite{Ambjorn:1985iw,Ambjorn:1986fz} and also practical problems (runaway trajectories) appeared.
Recently, the method has enjoyed renewed attention and many of its
problems have been solved.
It has been proved that provided the action is holomorphic and
the distributions of the variables decay fast enough, the
complex Langevin equation will converge to the correct results
\cite{Aarts:2009uq,Aarts:2011ax,erhardplenarygranada}. In a recent study it has been proposed that
using a certain observable the magnitude of the 'boundary term' at infinity can
be estimated \cite{boundaryterms1}, with an observable that is cheap
to calculate also for lattice systems \cite{boundaryterms2}. 
 Gauge theories pose an additional problem: the complexification of the 
gauge degrees of freedom, which in turn leads to large fluctuations and 
the breakdown of the simulation.  
The procedure of gauge cooling \cite{gaugecooling,Aarts:2013uxa} 
 was introduced to mitigate this problem.
With the help of gauge cooling it became possible to simulate 
HDQCD (heavy dense QCD) where the quarks are kept static  
\cite{gaugecooling,Aarts:2013uxa}, as well as full QCD using light
quarks in the 
staggered \cite{Sexty:2013ica,Schmalzbauer:2016pbg,Bloch:2017jzi,Nagata:2018mkb,Tsutsui:2018jva,Kogut:2019qmi} and the Wilson
formulation \cite{Aarts:2014bwa}, 
as well as QCD with a theta term \cite{Bongiovanni:2014rna}.

Reaching is the continuum limit in lattice calculations is a non-trivial
task. The usual strategy is to use improved actions which are more
expensive numerically but they ensure a quicker convergence.
Using an improved gauge action with CLE is straightforward,
but fermionic improvements can be more involved.
In this paper the stout smearing procedure\cite{Morningstar:2003gk} is
generalized to ensure applicability in the Complex Langevin setup.

In Section \ref{clesec} a brief overview of the 
complex Langevin method is given.
In Section \ref{pressec} the strategy of the calculation of the pressure and
energy density is discussed. Afterwards, the stout smearing procedure is
generalized to the complexified manifold of the link variables in
Section \ref{stoutsec}.
In Section \ref{ressec} the numerical results are presented.
Finally, conclusions are offered in Section \ref{concsec}.

\section{The complex Langevin equation}
\label{clesec}

The Langevin equation for $U_{x,\nu} \in$ SU($N$), 
the link variables of a gauge theory, in discretised  
form with Langevin timestep $\epsilon$  
is written as \cite{PhysRevD.32.2736}:

\bea
U_{x,\nu} (\tau+\epsilon) = 
  \textrm{exp} \left[ i \sum\limits_a 
 \lambda_a ( \epsilon K_{ax\nu}  + \sqrt\epsilon \eta_{ax\nu} ) 
\right]         U_{x,\nu}(\tau),
\eea
with $\lambda_a$ the generators of the gauge group, i.e. 
the Gell-Mann matrices, the drift force $K_{ax\nu} = -(D_{ax\nu} \rho[U])/\rho[U] $ 
calculated from the measure $\rho$ using the left derivative
\bea
D_{ax\nu}f(U) = \left. \partial_\alpha f( e^{i \alpha \lambda_a} U_{x,\nu}) \right|_{\alpha=0}.
\eea
For a complex measure the drift
terms become complex with $K_{ax\nu} \in \mathcal{C}$. 
The manifold of the link variables is then complexified to SL(3,$\mathcal{C}$). 
In the case of lattice QCD with fermions the action is 
written as
\bea
S_{eff}= S_g - \textrm{ln det} M(\mu),
\eea
with the determinant of the fermionic Dirac matrix $M(\mu)$. The measure
$ e^{-S_{eff}} = e^{-S_g} \textrm{det} M(\mu) $ generally
has zeroes on the complexified field manifold, resulting in meromorphic
drift terms.
Simulating such a theory with the CLE can potentially lead to
incorrect results. It has been shown that in the case of QCD
at large temperatures these zeroes are not reached by the process,
thus the formal justification of the Complex Langevin method
goes through \cite{Aarts:2017vrv}.

The non-unitarity of the link variables can be monitored using the 
unitarity norm
\bea
\label{unitaritynorm}
{ 1 \over 4 \Omega} \sum_{x,\mu} \Tr (( U_{x,\mu} U^{+}_{x,\mu} -1 )^2),
\eea
where $\Omega = N^3_s N_t $ is the space-time volume of the lattice.
The uncontrolled growth of the unitarity norm observed in naive 
complex Langevin simulations can be countered using 
complexified gauge transformations after each update such that 
the unitarity norm is decreased, i.e. gauge cooling 
\cite{gaugecooling,Aarts:2013uxa}
(see also \cite{Nagata:2015uga} for the inclusion of gauge cooling into
the formal proof of correctness).
It has been observed that 
gauge cooling is effective as long as the $\beta$ parameter of the theory
is not too small \cite{Aarts:2013nja}. The minimal $\beta$ corresponds to 
a maximal lattice spacing, such that the continuum limit 
can be carried out in the safe region, allowing the mapping of the 
phase diagram of the HDQCD theory \cite{hdqcdpd}.

In this study the naive plaquette and staggered action is
used as well as the tree-level Symanzik improved gauge action with stout
smeared staggered fermions\cite{Morningstar:2003gk}.
The implementation of the gauge actions is straightforward: one
ensures the holomorphicity of the action by replacing the
matrix adjungate with the matrix inverse for the plaquette and extended
plaquette variables appearing in the action. 
The naive fermionic drift is calculated with the help of
noise vectors \cite{Sexty:2013ica}, 
and the implementation of the stout smearing is detailed in Sec.~\ref{stoutsec}.

\section{Thermodynamics at nonzero chemical potential}
\label{pressec}

Using the grand canonical ensemble the pressure in units of $T^4$ is calculated from the grand partition function $ \mathcal{Z}(T,\mu) $:
\bea
 { p \over T^4} = {\ln \mathcal{Z} \over V T^3 }
\eea

For the purposes of this study we assume that the pressure
calculation at zero chemical potential has been carried out by some method. Our primary interest here is the change
of the pressure as
the chemical potential is increased at a fixed temperature,
since a direct calculation
at $ \mu>0$ is not possible with the usual importance
sampling simulations
\bea
\Delta\left(p \over T^4 \right) = \left.   p \over T^4  \right|_{T,\mu} -  \left. p \over T^4 \right|_{T,\mu=0}.
\label{eq:pressurediff}
\eea
Usually one Taylor-expands this difference at $\mu=0$ to allow
calculations of the coefficients using Monte-Carlo simulations \cite{Allton:2003vx,Allton:2005gk}
\bea
\Delta\left(p \over T^4 \right) = \sum_{n>0, even} c_n(T) \left( \mu \over T \right)^n,
\eea
with
\bea
\label{cndef}
c_n(T)
= {1\over n!} { N_t^3\over N_s^3} \left. { \partial^n \ln \mathcal{Z} \over \partial (\mu N_t)^n } \right|_{\mu=0}.
\eea
From the symmetry of the partition function $ Z(\mu)=Z(-\mu)$ we see that
only the even coefficients $ c_{2k}$ are nonzero. The derivatives
in $c_n$ can be expressed as expectation values of traces
of operators involving $M^{-1}$ and $\partial_\mu M$, measured
at $\mu=0$. For example, $c_2$ is evaluated using 
\bea
    {\partial^2 \ln \mathcal{Z} \over \partial \mu^2}
    = \left\langle {N_F\over 4} { \partial^2 \ln \det M \over \partial \mu^2 }
    \right \rangle
    + \left\langle \left( {N_F\over 4} { \partial \ln\det M \over \partial \mu}
     \right)^2 \right\rangle,
     \eea
     where the derivatives of $ \ln \det M $ are given by
\bea
    { \partial \ln\det M \over \partial \mu}
    &=& \Tr \left( M^{-1} \partial_\mu M \right)
  \\ \nonumber 
 { \partial^2 \ln \det M \over \partial \mu^2 }
 &=& \Tr \left( M^{-1} \partial^2_\mu M \right) -
 \Tr \left( M^{-1} (\partial_\mu M)  M^{-1} \partial_\mu M \right).
\eea
Higher derivatives involving more and more terms and higher powers
of $M^{-1}$ and $\partial_\mu M$ can be found in e.g. \cite{Allton:2005gk}.

Using the Complex Langevin equation we can simulate at nonzero
chemical potential so the pressure is accessible as:

\bea \label{pressureint}
\Delta\left(p \over T^4 \right) &=&
           { \ln \mathcal{Z}(T,\mu) - \ln \mathcal{Z}(T,0) \over V T^3 } =
           {1 \over VT^3} \int_0^\mu d\mu' {\partial\ln \mathcal{Z}(T,\mu') \over \partial \mu'} 
           = {1 \over VT^3} \int_0^\mu d\mu' \Omega n(\mu'),     
\eea
where we have defined the
charge density (using the space-time volume $\Omega=N_s^3 N_t$)
\bea 
n = {1\over \Omega} {\partial \ln \mathcal{Z} \over \partial \mu }
=  {N_f \over 4 \Omega} \big\langle \Tr ( M^{-1} \partial_\mu M ) \big\rangle
\eea
This means we can calculate the pressure at high chemical potentials
at the cost of measuring the density at several chemical potentials
in between and performing the integral (\ref{pressureint}).
The density is a cheap observable with relatively small fluctuations.
In contrast, for the Taylor expansion one needs to measure
the $c_n$ coefficients at $\mu=0$, however these are quite costly,
as they involve many inversions as $n$ increases, and they tend to
be very noisy with increasing $n$, such that state of the art
calculations
can measure coefficients up to $c_6$ with a great
effort \cite{Bazavov:2017dus}, but also other
approaches exist based on imaginary chemical
potentials \cite{Delia:2016jqh,Borsanyi:2018grb}. The extrapolated
results to $\mu>0$ have error bars increasing such that
they quickly lose predictive power above $ \mu/T \sim 1$.

Once the pressure is calculated, i.e. the grand partition function
is reconstructed, other thermodynamical observables can be calculated from
it using various derivatives. The $\mu$ dependence of the density and
fluctuations of the density can be directly measured in a simulation
at the $\mu$ value of interest.
Below the calculation of the energy density
is detailed, the calculation of further quantities such as entropy density,
speed of sound, charge susceptibilities, etc. is beyond the
scope of this study.

The energy density $\epsilon$ can be accessed from the grand partition
function through the trace anomaly
\bea
 \label{traceanomaly}
    {\epsilon - 3 p \over T^4 } = - {1 \over V T^3}
    a \left({ \partial \beta \over \partial a }\right)_\textrm{LCP}
    \left[ { \partial \textrm{ln} \mathcal{Z} \over \partial \beta } + \left({ \partial m \over \partial \beta }\right)_\textrm{LCP} { \partial \textrm{ln} 
        \mathcal{Z} \over \partial m } \right],
\eea
where $\beta$ and $m$ are the bare parameters of the action, and this
formula is also valid at $\mu>0$ \cite{Allton:2003vx}. As indicated, the derivatives
in the formula above are understood to be defined along
the line of constant physics (LCP), where the pion mass is
kept fixed in physical units.
In the importance sampling formulation eq.(\ref{traceanomaly}) is Taylor
expanded in $\mu$ similarly
to the pressure.
For the first nonzero coefficient of the Taylor expansion (at the second order)
one then measures the observables 
\bea
{\partial^3 \textrm{ln} \mathcal{Z} \over \partial \beta \partial \mu^2 }
,\ \ \ \  
{\partial^3 \textrm{ln} \mathcal{Z} \over \partial m \partial \mu^2 },
\eea
using
\bea
    { \partial \langle O \rangle \over
      \partial \beta } = -\left\langle O {\partial S_g \over \partial \beta }
    \right\rangle 
    + \langle O \rangle \langle {\partial S_g \over \partial \beta } \rangle, 
    \\ \nonumber
{ \partial \langle O \rangle \over
  \partial m }
= \left \langle {\partial O \over \partial m } \right\rangle
+ \Omega \left \langle O \chi \right \rangle 
-\Omega  \langle O \rangle \langle  \chi  \rangle
\eea
with the chiral condensate $ \chi = (N_F/4\Omega)         \partial \textrm{ln} \textrm{det} M /\partial m  $.
In the complex Langevin setup however the $\mu$ dependence
\bea
\Delta \left( {\epsilon -3p \over T^4 } \right) =
\left. {\epsilon  -3p \over T^4 }\right|_{\mu} -
\left. {\epsilon  -3p \over T^4 }\right|_{\mu=0}
\eea
can be directly calculated using two simulations at the $\mu$
value of interest and at $\mu=0$. The observables needed for this calculation
are the gauge action average $ \langle S_g \rangle 
= -\partial \textrm{ln} \mathcal{Z} / \partial {\beta } $
and the chiral condensate
$\chi = ( \partial \textrm{ln} \mathcal{Z} / \partial m ) T/V$.
The beta function $ a ( \partial \beta / \partial a )_\textrm{LCP} $
and the derivative $  \left({ \partial m / \partial \beta }\right)_\textrm{LCP} $
can be estimated by independent simulations
at zero temperature and $\mu=0$.

\section{Stout smearing}
\label{stoutsec}

To use stout smearing in Complex Langevin simulations, we must generalize
its domain of definition from SU(N) to SL(3,$\mathcal{C}$) matrices,
using a holomorphic function. The weighted staple sum corresponding
to a link variable $U_\nu(x)$ is given by 
\bea
C_\nu(x) = \sum_{\sigma\neq\nu} \rho_{\nu\sigma} \left(
 U_\sigma(x) U_\nu(x+\hat\sigma) U_\sigma^{-1}(x+\hat\nu)
+ U^{-1}_\sigma(x-\hat\sigma) U_\nu(x-\hat\sigma) U_\sigma(x-\hat\sigma+\hat\nu)
\right),
\eea
where $\rho_{\nu\sigma}$ are some real weights.
Since we cannot use adjungation, we also need the sum of the inverse paths:
\bea
Z_\nu(x) = \sum_{\sigma\neq\nu} \rho_{\nu\sigma} \left(
 U_\sigma(x+\hat\nu) U^{-1}_\nu(x+\hat\sigma)  U^{-1}_\sigma(x)
+ U^{-1}_\sigma(x-\hat\sigma+\hat\nu)
 U^{-1}_\nu(x-\hat\sigma)  U_\sigma(x-\hat\sigma)
\right).
\eea

We than define
\bea \Omega_\nu(x)& =& C_\nu(x) U^{-1}_\nu(x), \\ \nonumber
  \Omega^i_\nu(x)& =& U_\nu(x) Z_\nu(x), \\ \nonumber
 X_\nu(x) = iQ_\nu(x) &=& {1\over 2} ( \Omega_\nu(x)  - \Omega^i_\nu(x) ) - {1 \over 2 N}
 \Tr  ( \Omega_\nu(x)  - \Omega^i_\nu(x) )
 \eea
 and finally the smeared field is given by $ U_\nu'(x)= e^{iQ_\nu(x)} U_\nu(x) $.
 This definition coincides with the usual stout smearing if the gauge fields
 are in SU(N), and the matrix $ Q_\nu(x)$ is Hermitian in this case.
 On SL(3,$\mathcal{C}$) $ Q_\nu(x)$ is no longer Hermitian but it is still
 traceless,
 so $U'_\nu(x)$ is also an element of SL(N,$\mathcal{C}$). Typically
 one takes multiple smearing steps with
 \bea
 U \rightarrow U^{(1)} \rightarrow ... \rightarrow U^{(n)}
 \eea
 and the measure becomes $ e^{-S(U)} = e^{-S_g(U)} \textrm{det} \left( M( U^{(n)}) \right)$, where $S_g(U)$ is the gauge action and $M(U)$ is the Dirac matrix describing the fermionic degrees of freedom.

 For the calculation of the drift terms we need to evaluate
 $D_{a\nu x} S(U)$. Since the gauge part does not involve smearing we
 write $ S(U)= S_g(U)+ S_f(U) $ and we only consider the fermionic drift
 $D_{a\nu x} S_f(U)$ below.

 Let's consider one smearing step first where
 $ U'_\nu=e^{X_\nu} U_\nu$, and $ i \lambda_a  D'_{a\sigma} S_f(U') =F'_\sigma$ is the standard force
 for unimproved fermions (with $D'_{a\sigma}$ the left derivative with respect to variable $U'_\sigma$), and the space-time indices  are suppressed. 
 Our aim is to calculate $ F_\nu = i \lambda_a D_{a\nu} S_f(U) $, the force
 of the unsmeared
 field. For multiple smearing steps the procedure detailed below
 is repeated iteratively.
  For the drift term we will need to evaluate the left 
 derivative  $ D_{a\nu} U'_\sigma$, which can be represented as 
 \bea
D_{ab\nu\sigma}= - {i\over 2} \textrm{Tr} \left( \lambda_b  (D_{a\nu} U'_\sigma) U'^{-1}_\sigma \right),
 \eea
 such that to first order in $\alpha_a$
 \bea
U'_\sigma(e^{i \alpha_a \lambda_a } U_\nu) = e^{i \alpha_a D_{ab\nu\sigma} \lambda_b } U'_\sigma,
 \eea
 and the chain rule is written as $ D_{a\nu} S[U]= D_{ab\nu\sigma} D_{b\sigma}'S[U'] $.
 The drift term is then written (also using the product rule $ D_a (e^X U) =  (D_a e^X) U  + e^X D_a U $ and the identity $ -(i \lambda_a /2) \Tr (i \lambda_a W ) = W - (1/N) \Tr W $)   
 \bea
 F_\nu = i \lambda_a  D_{a\nu} S[U] = - { i \lambda_a  \over 2 } \Tr \left( F'_\sigma (D_{a\nu} U'_\sigma) U'^{-1}_\sigma \right)= - { i \lambda_a \over 2 }
 \Tr \left ( e^{-X_\sigma} F'_\sigma D_{a\nu} e^{X_\sigma} \right) + e^{-X_\nu} F'_\nu e^{X_\nu} - {1\over N} \Tr \left( e^{-X_\nu} F'_\nu e^{X_\nu} \right)  
 \eea
 We write $ \Tr (e^{-X_\sigma} F'_\sigma  D_{a\nu} e^{X_\sigma}) = \Tr ( L_\sigma(X,F') D_{a\nu} X_\sigma ) $ where one can take $L_\sigma(X,F')$ to be traceless (and anti-Hermitian for unitary link variables).
Using the definition of $X$ we obtain (for isotropic smearing with $\rho_{\nu\sigma}=\rho $):
 \bea
 F_\nu(x) &=& i \lambda_a D_{a\nu x} S =
  \Bigg[ e^{-X_\nu(x)} F'_\nu(x)  e^{X_\nu(x)}
 - {1 \over 2 } \left( L_\nu(x)  \Omega_\nu(x) + \Omega^i_\nu(x) L_\nu(x) \right)   \\ \nonumber
 &&+ {\rho\over 2} \sum_{\sigma\neq \nu}
 \Big( U_\nu(x) U_\sigma(x+\nu) U^{-1}_\nu(x+\sigma) U^{-1}_\sigma(x) L_\sigma(x)
  + L_\sigma(x)  U_\sigma(x) U_\nu(x+\sigma) U^{-1}_\sigma(x+\nu) U^{-1}_\nu(x)
\\ \nonumber
&& + U_\nu(x) U^{-1}_\sigma(x+\nu-\sigma) U^{-1}_\nu(x-\sigma)  L_\nu(x-\sigma) U_\sigma(x-\sigma) 
  + U^{-1}_\sigma(x-\sigma)  L_\nu(x-\sigma) U_\nu(x-\sigma)  U_\sigma(x+\nu-\sigma) U^{-1}_\nu(x) 
\\ \nonumber
&&  - U^{-1}_\sigma(x-\sigma)  L_\sigma(x-\sigma) U_\nu(x-\sigma) U_\sigma(x+\nu-\sigma) U^{-1}_\nu(x)
- U_\nu(x) U^{-1}_\sigma(x+\nu-\sigma) U^{-1}_\nu(x-\sigma ) L_\sigma(x-\sigma) U_\sigma(x-\sigma) \\ \nonumber
&& - U_\sigma(x) U_\nu(x+\sigma) U^{-1}_\sigma(x+\nu) L_\sigma(x+\nu) U^{-1}_\nu(x)
   -   U_\nu(x) L_\sigma(x+\nu) U_\sigma(x+\nu) U^{-1}_\nu(x+\sigma) U^{-1}_\sigma(x)
   \\ \nonumber
   &&
   + U_\nu(x) U_\sigma(x+\nu) U^{-1}_\nu(x+\sigma) L_\nu(x+\sigma) U^{-1}_\sigma(x)
   +   U_\sigma(x) L_\nu(x+\sigma) U_\nu(x+\sigma) U^{-1}_\sigma(x+\nu) U^{-1}_\nu(x)
   \\ \nonumber
   && 
   + U_\nu(x) U^{-1}_\sigma (x+\nu-\sigma)L_\sigma(x+\mu-\sigma) U^{-1}_\nu(x-\sigma) U_\sigma(x-\sigma)
   \\ \nonumber
 &&  
   + U^{-1}_\sigma(x-\sigma) U_\nu(x-\sigma) L_\sigma(x+\mu-\sigma) U_\sigma(x+\mu-\sigma) U^{-1}_\nu(x) \Big)
  \Bigg]_\textrm{traceless part}.
 \eea

 Finally, to calculate the matrix $L_\sigma(X,F')$ one can proceed using the following theorem \cite{expderiv}:
 For a matrix $X$ of size $ N \times N $, we write $ \exp (t X ) $ as
 \bea
 \exp ( t X ) =  V \textrm{diag}(\exp (t \lambda_0), ...., \exp
 (t \lambda_n)) V^{-1} 
\eea
Where $\lambda_i$ are the eigenvalues and $V$ is the matrix whose $j$th column
is the eigenvector of $\lambda_j$.
We than have
\bea
{\partial e^{tX} \over \partial \theta } =  V ( G \times E ) V^{-1},
\eea
where the cross-product is defined as $ ( G \times E )_{ij} = G_{ij} E_{ij}$ (no summation), and $ G= V^{-1} (\partial X /\partial \theta ) V $. The matrix $E$ is defined as
\bea
E_{ij} & = & { (e^{t \lambda_i} - e^{t \lambda_j}) \over \lambda_i - \lambda_j} 
\qquad \textrm{for }   i \neq j \\ \nonumber
E_{ii} & = & t e^ { t \lambda_i } \qquad  \qquad \qquad \textrm{for }  i = j.
\eea
This leads to $ \Tr( R  D_a e^X ) = \Tr (   V  (  (V^{-1} R V)  \times E ) V^{-1}   D_a X )$.
Alternatively, using the Cayley-Hamilton theorem any analytical function of a traceless $ 3 \times 3$ matrix can be written as:
\bea
f(X)= f_0 + f_1 X + f_2 X^2,
\eea
where $f_i$ depends on the invariants of the matrix, $c_0=\textrm{det} X=\Tr (X^3)/3 ,\  c_1 = \Tr (X^2)/2$ (recall that $\Tr X=0$).
Consequently the derivative is written as:
\bea
D_a f(X) = D_a f_0 + D_a f_1 X + D_a f_2 X^2 + f_1 D_a X + f_2 ( (D_aX) X + X D_a X), \\ \nonumber
D_a f_i = { \partial f_i \over \partial c_0 } \Tr ( X^2 D_a X)
+ { \partial f_i \over \partial c_1 } \Tr ( X D_a X).
\eea
Calculating the coefficients $f_i$ and their derivatives for the exponential
function needed here proceeds by using
a polynomial approximation to a fixed order ensuring correct results up to
machine precision. Finally we write $ \Tr( R D_a e^X ) = \Tr ( B D_a X) $, with
\bea
B=
\Tr \left( R \sum_{i=0}^2 { \partial f_i \over \partial c_0 } X^i \right) X^2
+\Tr \left( R \sum_{i=0}^2 { \partial f_i \over \partial c_1 } X^i \right) X
+ f_1 R + f_2 ( RX+XR) 
\eea

\begin{figure}
\begin{center}
\epsfig{file=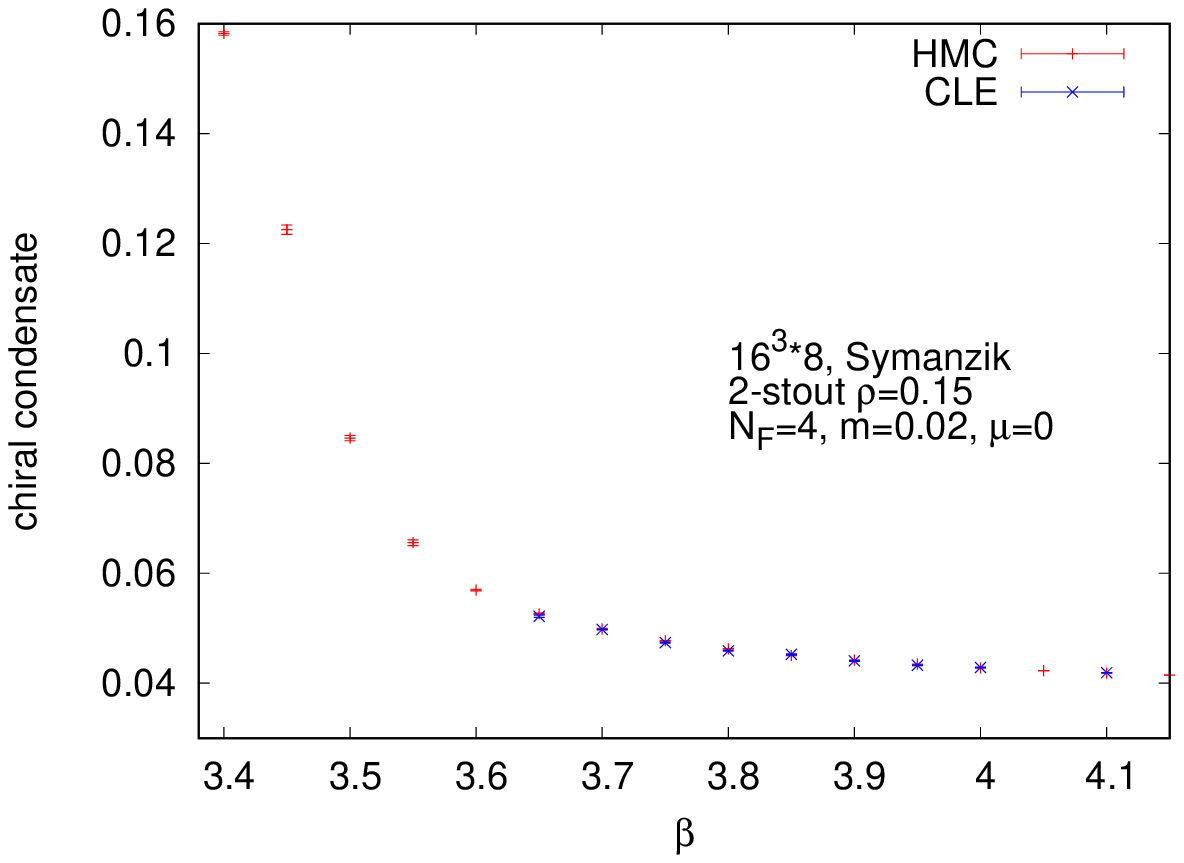, width=8.5cm}
\epsfig{file=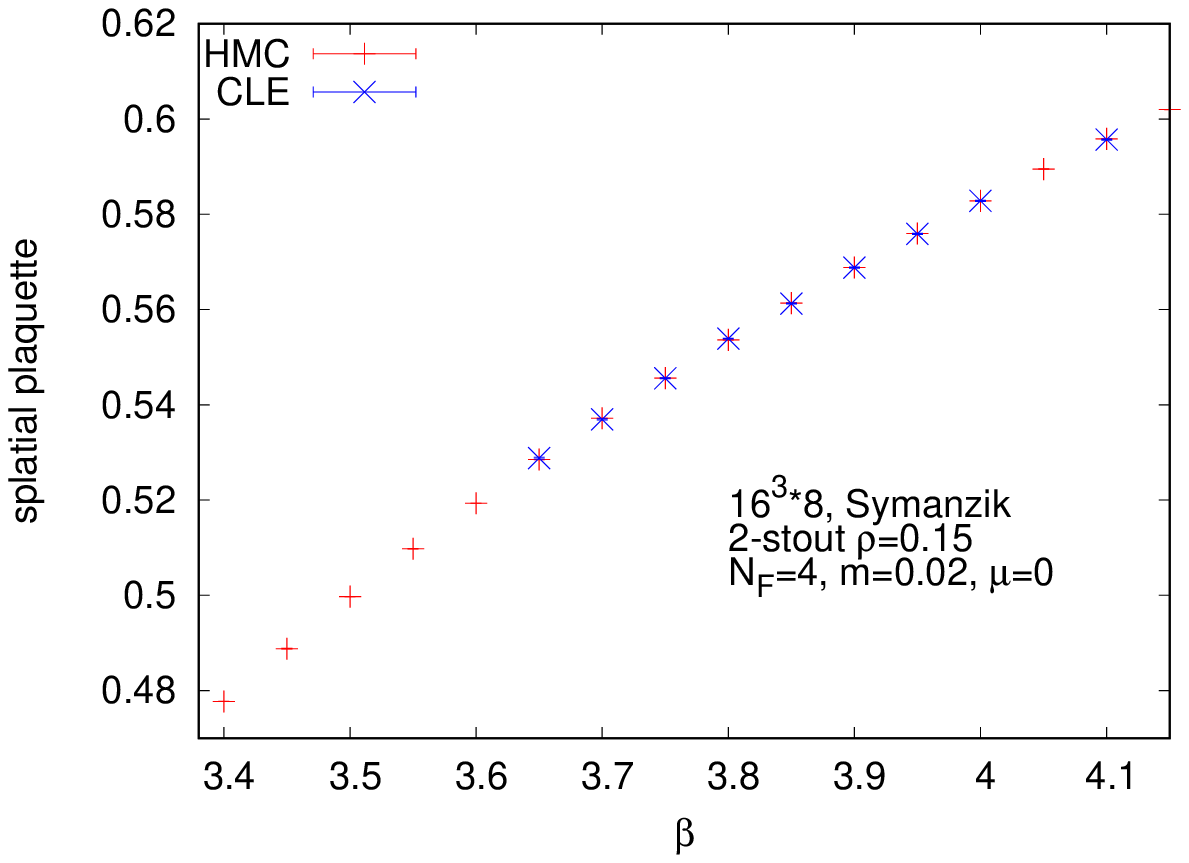, width=8.5cm}
\caption{Comparing HMC and CLE calculations with improved
action at $\mu=0$. }
\label{stoutbenchmark}
\end{center}
\end{figure}

To check that the implementation is correct I have benchmarked the CLE
results with results from the usual Hybrid Monte Carlo (HMC) implementation at $\mu=0$,
see in Fig.~\ref{stoutbenchmark}. The comparison used the
Symanzik gauge action and $n=2$ stout smeared staggered fermions with
$N_F=4$ and isotropic smearing with $\rho=0.15$. The simulations were started from an SU(3) configuration,
the observables are averaged between Langevin
times $10  < \tau <20 $. Agreement within statistical errorbars
is observed as long as the $\beta$ parameter is chosen large enough. For
smaller $\beta$ values the gauge cooling becomes less effective,
the unitarity norm rises quickly and the simulations
become instable, just as it was observed for the naive
action \cite{Fodor:2015doa}.

\begin{figure}[ht!]
\begin{center}
\epsfig{file=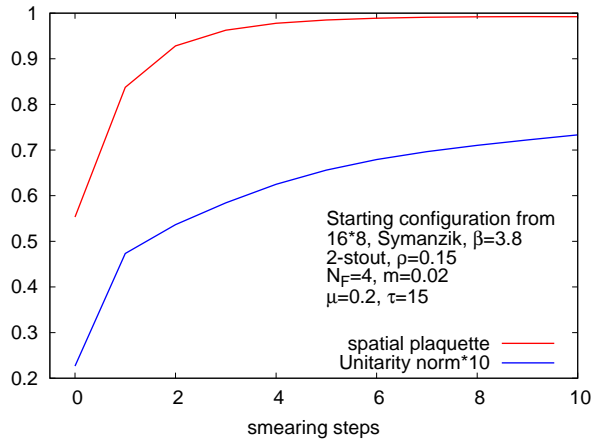, width=8.5cm}
\caption{The effect of the smearing on a typical configuration taken from a CLE simulation as indicated. Using isotropic smearing with $\rho=0.15$.}
\label{stout-behavior}
\end{center}
\end{figure}

In Fig.~\ref{stout-behavior} the effect of the smearing on a typical configuration from a CLE simulation is shown. The plaquette average nears 1.0 as it does also in the usual smearing of an SU(3) configuration. The unitarity norm of the configuration also increases slightly during the smearing procedure. If the initial unitarity norm of the configuration is higher $ \gtrsim 0.1 $, the smearing might cause a numerical overflow on the computer, especially if the $\rho$ parameter of the smearing is not small. This is similar to the 'runaway' behavior known to occur in complex Langevin simulations. For the simulations in this study parameters are chosen such that this breakdown is very unlikely to occur.

\section{Results}
\label{ressec}

Two actions used in this study, this gives a very rough estimate
of the cutoff effects, and it allows for the testing of the stout
staggered fermionic action with the Complex Langevin equation.
First I use the plaquette gauge action with the naive staggered formulation
using $N_F=4$ and the mass parameter $m=0.01$. Second the Symanzik gauge
action is used with a stout smeared staggered action using $n=2,\  \rho=0.15,\ 
N_F=4,\ m=0.02$. 
The lattice spacing (measured with the $w_0$ parameter \cite{Borsanyi:2012zs})
and the mass of the lightest pion taste is shown in Tables~\ref{tab:naiv},\ref{tab:stout}. 

\begin{table}[h!]
  \begin{tabular}{|c|c|c|c|c|}
    \hline
  $\beta$ & $a$(fm) & $m_\pi a$ & $T$(MeV) for $N_T=8$ & pion mass(MeV) \\
\hline
5 & $ 0.2892 \pm 0.0002 $ & $ 0.2595 \pm 0.0002 $ & 85.3 & 177  \\ 
5.1 & $ 0.1895 \pm 0.0005 $ & $ 0.2881 \pm 0.0002 $ & 130 & 300  \\ 
5.2 & $ 0.1105 \pm 0.0004 $ & $ 0.2965 \pm 0.0004 $ & 223 & 529  \\ 
5.3 & $ 0.0822 \pm 0.0003 $ & $ 0.2727 \pm 0.0005 $ & 300 & 654  \\ 
5.4 & $ 0.0633 \pm 0.0005 $ & $ 0.2496 \pm 0.0016 $ & 389 & 777  \\ 
5.5 & $ 0.0503 \pm 0.0005 $ & $ 0.2253 \pm 0.0016 $ & 490 & 883  \\ 
5.6 & $ 0.0433 \pm 0.0006 $ & $ 0.2229 \pm 0.0015 $ & 570 & 1020  \\
\hline
\end{tabular}
  \caption{The lattice spacing and the pion mass using the plaquette action
    with naive staggered fermions with $N_F=4,\ m=0.01,$ measured on a $24^3\times48$ lattice.}
\label{tab:naiv}
\end{table}

\begin{table}[h!]
\begin{tabular}{|c|c|c|c|c|}
  \hline
  $\beta$ & $a$(fm) & $m_\pi a$ & $T$(MeV) for $N_T=8$ & pion mass(MeV) \\
  \hline
3.5 & $ 0.1474 \pm 0.0004 $ & $ 0.3111 \pm 0.0004 $ & 167 & 417  \\ 
3.6 & $ 0.1159 \pm 0.0003 $ & $ 0.2790 \pm 0.0004 $ & 213 & 475  \\ 
3.7 & $ 0.0946 \pm 0.0005 $ & $ 0.2515 \pm 0.0005 $ & 261 & 525  \\ 
3.8 & $ 0.0769 \pm 0.0004 $ & $ 0.2259 \pm 0.0009 $ & 321 & 579  \\ 
3.9 & $ 0.0644 \pm 0.0004 $ & $ 0.2088 \pm 0.0016 $ & 383 & 640  \\ 
4 & $ 0.0535 \pm 0.0004 $ & $ 0.1987 \pm 0.0024 $ & 461 & 733  \\ 
4.1 & $ 0.0415 \pm 0.0006 $ & $ 0.2119 \pm 0.0052 $ & 594 & 1010  \\ 
  \hline
\end{tabular}
  \caption{The lattice spacing and the pion mass using the Symanzik gauge action
    with stout smeared staggered fermions with $N_F=4,\ m=0.02,\ n=2,\ \rho=0.15,$ measured on a $24^3\times48$ lattice.}
\label{tab:stout}  
\end{table}

\begin{figure}
\begin{center}
  \epsfig{file=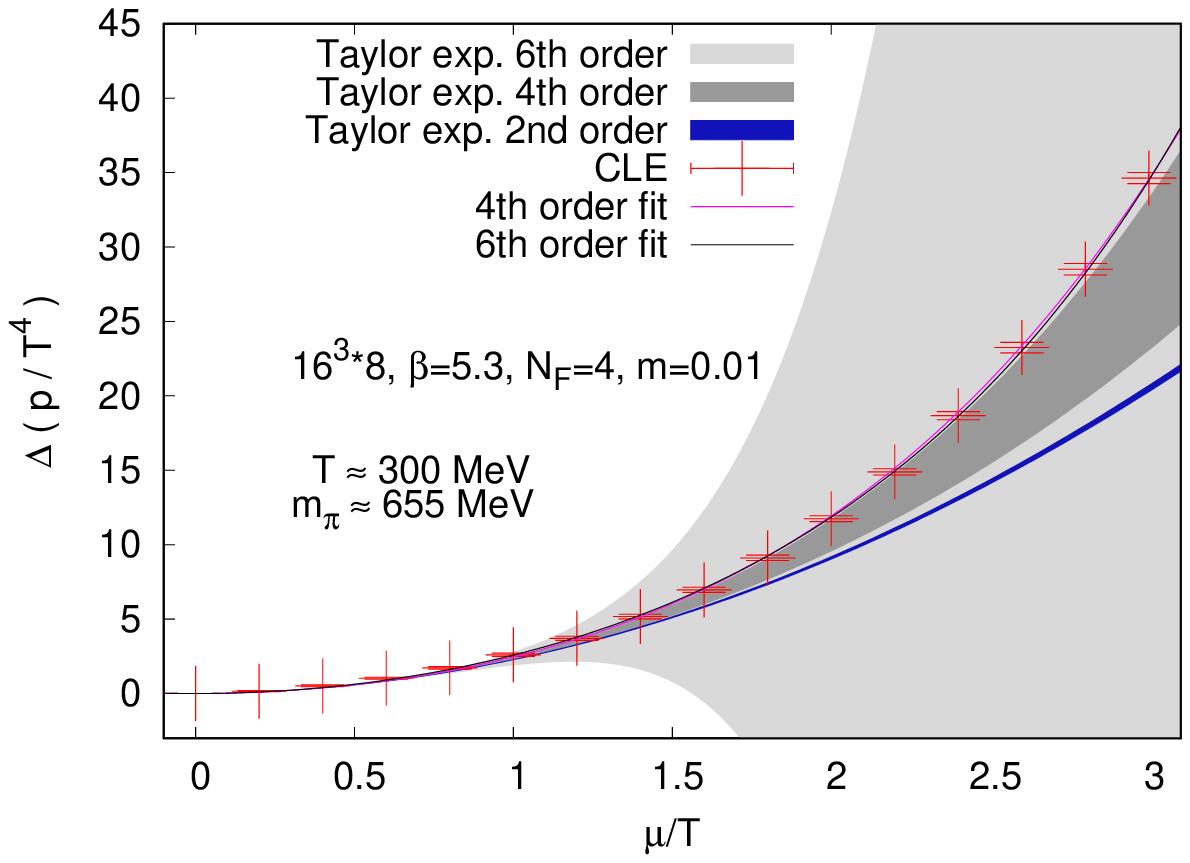, width=8.5cm}
  \epsfig{file=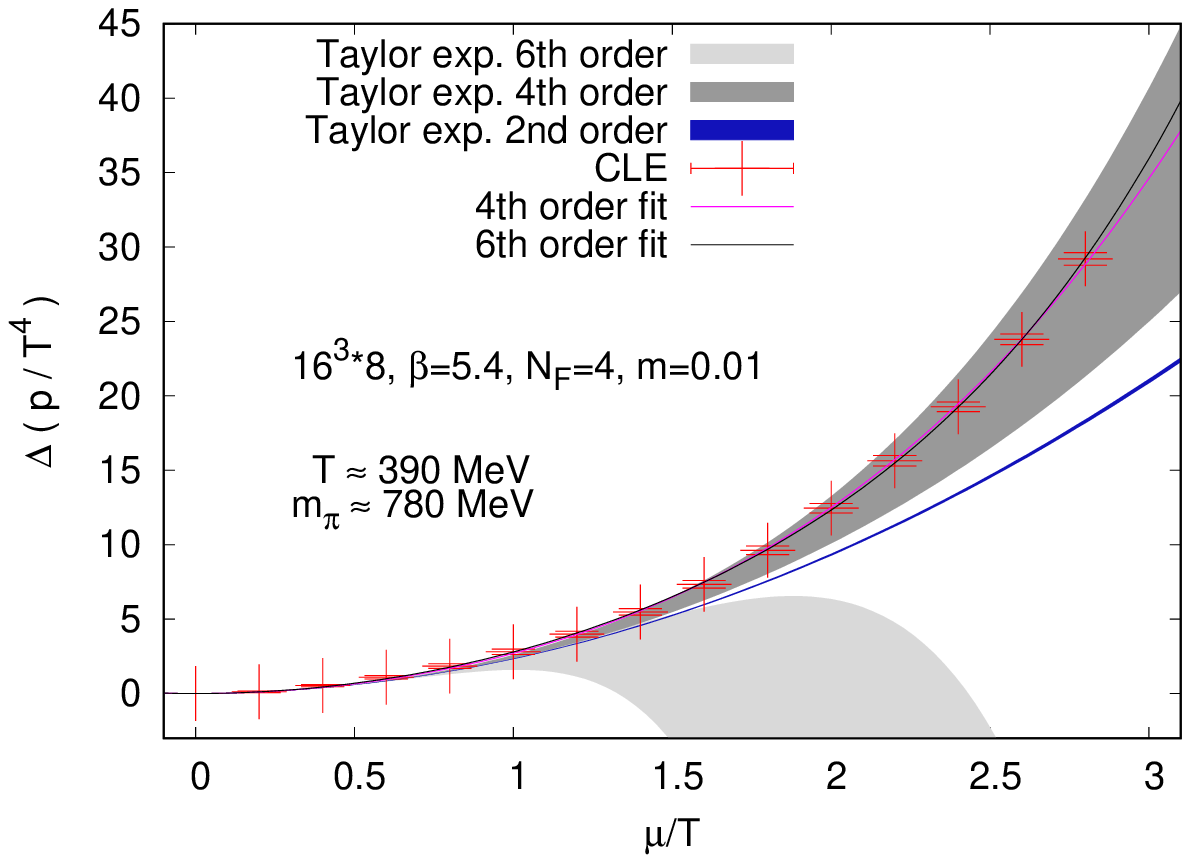, width=8.5cm}
\caption{
  The pressure difference defined in eq. (\ref{eq:pressurediff})
  for the naive action for two different lattice spacings.
}
\label{pressurefig}
\end{center}
\end{figure}

In Fig.~\ref{pressurefig} the pressure difference (\ref{eq:pressurediff})
is shown
for the naive ensemble for two different lattice spacings. To estimate the
Taylor coefficients, $\approx 1000$ configurations were generated using
a HMC simulation and on each configuration
the $c_n$ were estimated using 128 noise vectors.
The temperature
of the system is above the deconfinement transition for both lattice spacings. The Taylor coefficients
are listed in Table~\ref{tab:naivcn}.
Note that while in the continuum limit the Stefan-Boltzmann(SB) limit of $c_2$
is $N_F/2=2$, in the $ N_t=8$ discretisation the SB limit is expected
to be $\approx 2.8$ \cite{Allton:2003vx}.
To apply the integration method (\ref{pressureint}) the integral is discretised
with the stepsize $ a \Delta \mu = 0.025$, and CLE simulations are carried out
at each chemical potential.
The simulations used a partially second order
update scheme \cite{Fukugita:1986tg}
with adaptive control of the Langevin stepsize \cite{Aarts:2009dg},
using control parameters such that the timestep was
typically in the range $ (0.5- 1 )10^{-4}$.
The simulations are started from a configuration
where the link variables are initialized with white noise
in SU(3) directions only.
The thermalization of physical quantities such as the plaquette average, Polyakov loop average, density, etc. follows the expected exponential relaxation  
$ \sim e^{-\tau/\tau_0}$ with $\tau_0<1$ for all parameters.
3 runs are used to collect averages for
Langevin times $ 10 < \tau < 20 $.
The pressure is then reconstructed numerically
and statistical errors are estimated using the jackknife method
by splitting the stream of measurements to 10 pieces.
Since the density as a function of the chemical potential is reasonably
smooth at the high temperatures employed here,
the systematic error coming from the discretisation of this
integral is small (smaller than the statistical errors in this case), as can
be estimated by employing different $ \Delta \mu $ stepsizes.
Quark chemical potentials up to $ \mu= 4T$ are used, 
this corresponds to $\mu a =0.5$.
The Complex Langevin setup can be used for calculations at even higher
chemical potentials, however cutoff effects will become important
there.  One observes good agreement of the Taylor
expansion and the integration method. Note that  while the errorbars of the pressure calculated from the integration method are small, the estimation of the coefficients of the Taylor expansion includes the systematic error corresponding
to the choice of the fitting range.

\begin{table}[h!]
\begin{tabular}{|c|c|c|c|c|c|}
\hline
 $\beta$ & $c_2$ Taylor exp.  & $c_4$ Taylor exp. & $c_2$ CLE & $c_4$ CLE  \\
\hline
5.2 &  $ 2.102 \pm 0.059 $ & $ 0.233 \pm 0.17 $ & $2.21 \pm 0.2 $ & $0.15 \pm 0.05 $ \\
5.3 &  $ 2.277 \pm 0.026 $ & $ 0.095 \pm 0.06 $ & $2.24 \pm 0.1 $ & $0.18 \pm 0.05 $ \\
5.4 & $ 2.333 \pm 0.016 $ & $ 0.146 \pm 0.095 $ & $ 2.39 \pm 0.1$  & $0.16 \pm 0.02 $  \\
5.5 & $ 2.376 \pm 0.019 $ & $ 0.125 \pm 0.019 $ & $ 2.35 \pm 0.1$  & $0.18 \pm 0.02 $  \\
\hline
\end{tabular}
\caption{The coefficients of the Taylor expansion of the pressure 
  calculated at $\mu=0$ using eq. (\ref{cndef}) (label ``Taylor exp.'') and by fitting
  a polynomial to the results of the integration method (label ``CLE'').
  The unimproved action with $N_F=4,\ m=0.01$ is used
  on a $16^3\times 8$ lattice.
}
\label{tab:naivcn}
\end{table}

\begin{figure}[h!]
\begin{center}
  \epsfig{file=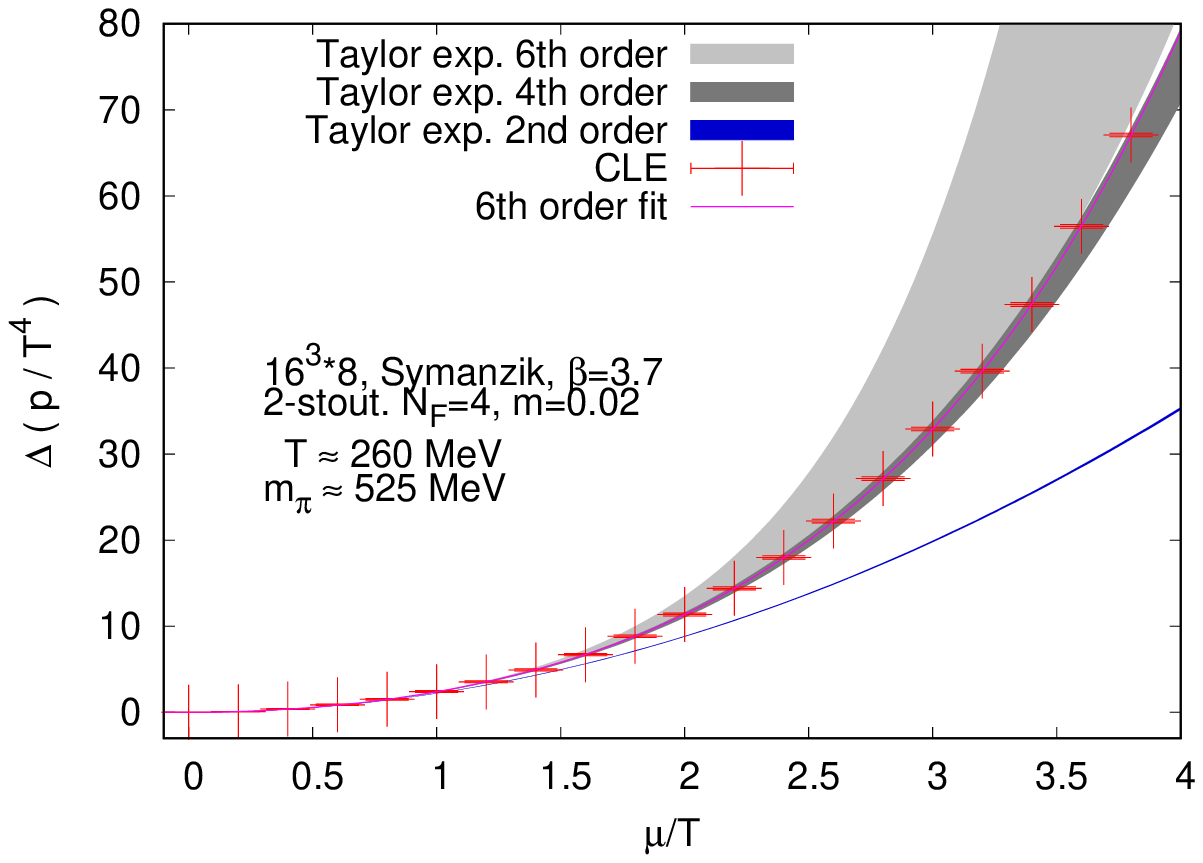, width=8.5cm}
  \epsfig{file=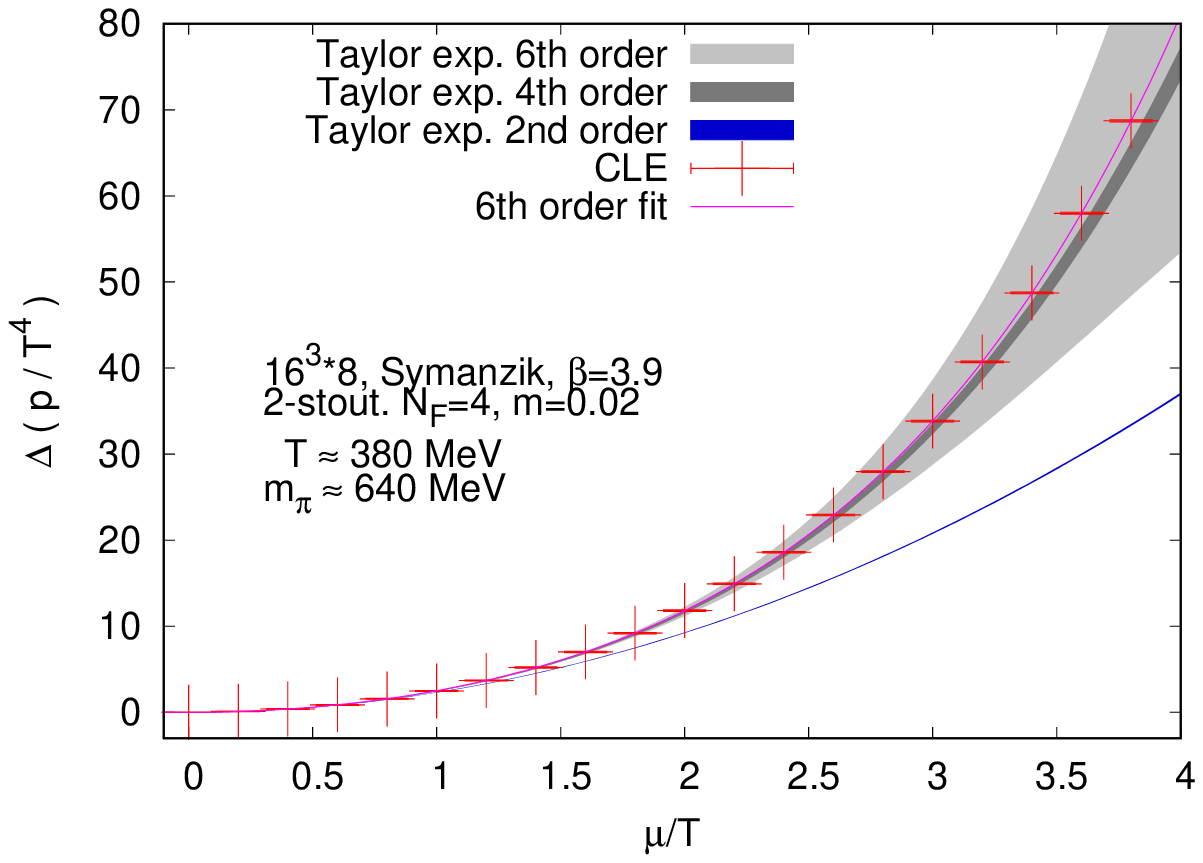, width=8.5cm}
\caption{
  The pressure difference defined in eq. (\ref{eq:pressurediff})
  for the improved action, using two different lattice spacings.
}
\label{pressurefig-stout}
\end{center}
\end{figure}

In Fig.~\ref{pressurefig-stout} the pressure difference is shown
for the improved ensemble for two lattice spacings.
The parameters were chosen such that the setup roughly corresponds
to the same physical lattice spacings and pion masses as the setup using
the unimproved action.
To measure the Taylor coefficients $\approx 2000$ configurations
from a HMC simulation were used with 64 noise vectors each.
The Taylor coefficients are listed in Table~\ref{tab:stoutcn}.
Using the improved action the importance sampling calculation of the $c_n$ coefficients
is slightly less noisy such that the $c_4$ is also calculated
with relatively small errors. The calculation of the $c_6$ coefficient
can also be attempted, however since it is quite small only an upper
limit on its magnitude is obtained.
One observes good agreement of the Taylor
expansion and the integration method, with relatively small discrepancy
of the CLE and 4th order Taylor expansion results
also at large $ \mu/T$, suggesting that
6th and higher order terms have very small coefficients.

\begin{table}[h!]
\begin{tabular}{|c|c|c|c|c|c|c|}
\hline
 $\beta$ & $c_2$ Taylor exp.  & $c_4$ Taylor exp. & $c_6$ Taylor exp. & $c_2$ CLE & $c_4$ CLE  & $c_6$ CLE \\
\hline
3.7 &  $ 2.206 \pm 0.009 $ & $ 0.156 \pm 0.016 $ & $ 0.016 \pm 0.013$ &
$2.33 \pm 0.1 $ & $0.13 \pm 0.02 $ & $0.002  \pm 0.001 $ \\
3.8 &  $ 2.293 \pm  0.007 $ & $ 0.171 \pm 0.017 $ & $ -0.01 \pm 0.01 $ &
$ 2.32 \pm 0.1 $ & $0.14 \pm 0.02 $ & $0.002 \pm 0.002 $ \\
3.9 & $ 2.312 \pm 0.007 $ & $ 0.150 \pm 0.007 $ & $ 0.001 \pm 0.005 $ &
$ 2.36 \pm 0.04 $  & $0.14 \pm 0.01 $  & $0.002 \pm 0.001 $ \\
4.0 & $ 2.371 \pm 0.012 $ & $ 0.124 \pm 0.009 $ & $ -0.001 \pm 0.006 $ &
$ 2.43  \pm 0.02$  & $0.13 \pm 0.01 $  & $0.002 \pm 0.001 $ \\
\hline
\end{tabular}
\caption{The coefficients of the Taylor expansion of the pressure 
  calculated at $\mu=0$ using eq. (\ref{cndef}) (label ``Taylor exp.'') and by fitting
  a polynomial to the results of the integration method (label ``CLE'').
  The 2-stout improved action is used with $ N_F=4, m=0.02, $ on a $16^3\times 8$ lattice.
}
\label{tab:stoutcn}
\end{table}

\begin{figure}[h!]
\begin{center}
  \epsfig{file=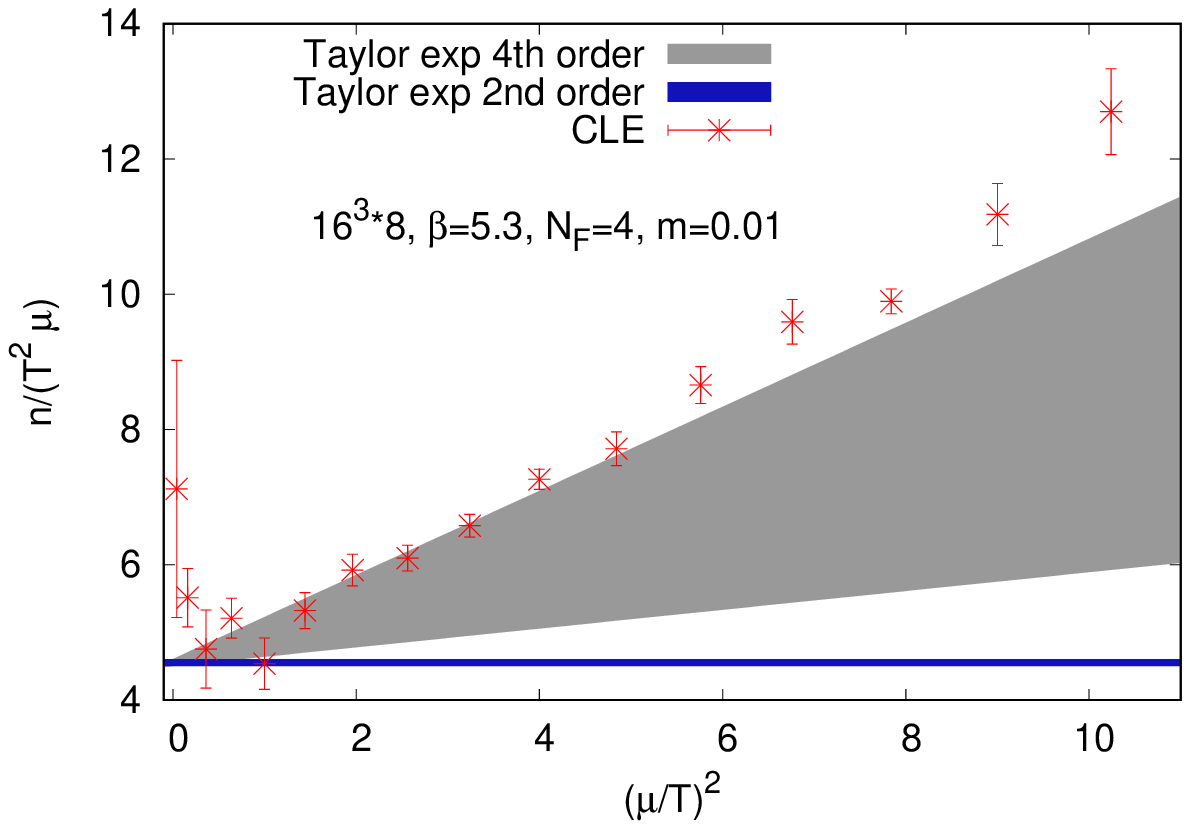, width=8.5cm}
  \epsfig{file=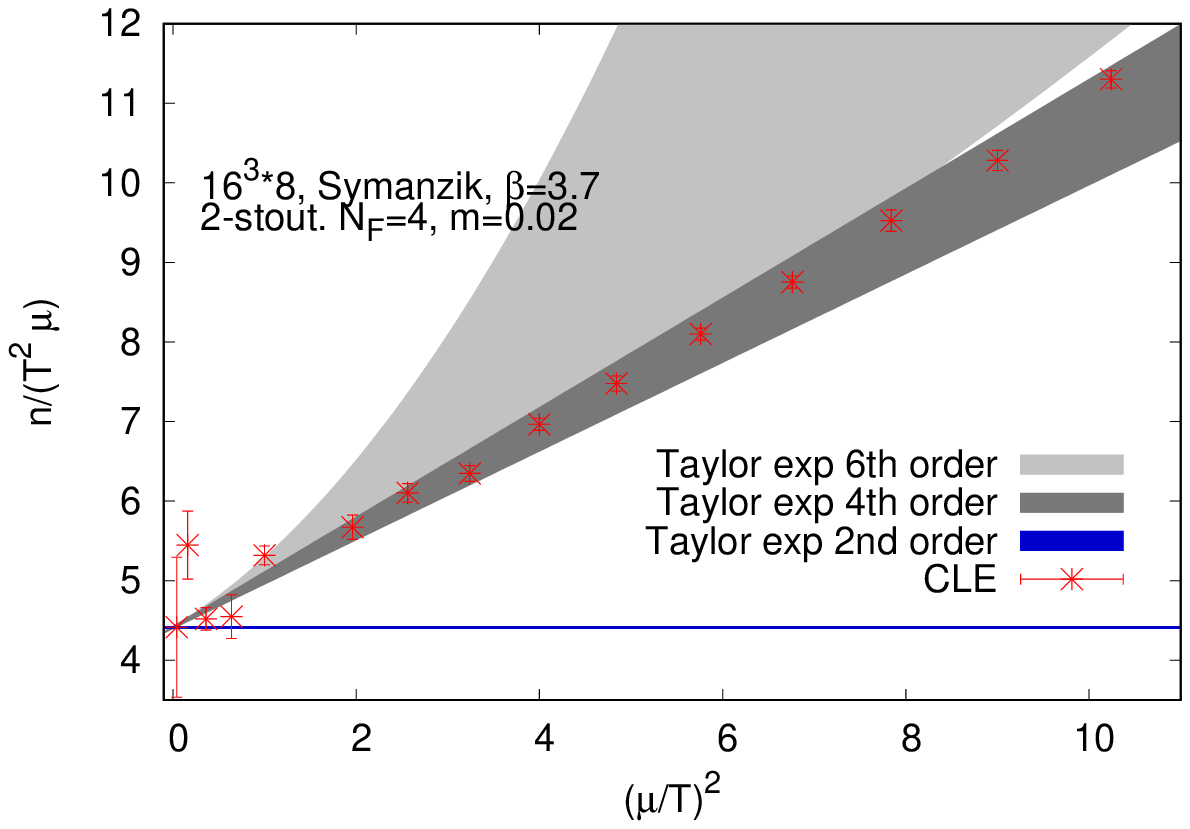, width=8.5cm}
\caption{
  $ n /(T^2 \mu) $ is plotted
  as a function of $ ( \mu/T)^2 $ for the naive action at $\beta=5.3$ (left)
  and the improved action at $\beta=3.7$ (right).
}
\label{linplot}
\end{center}
\end{figure}

In Fig.~\ref{linplot} the quantity
\bea
    {n \over T^2 \mu } = {T \over \mu } { \partial (p/T^4) \over \partial (\mu /T) }
\eea
is plotted as a function of $ ( \mu/T)^2 $.  This allows for an intuitive way
of judging the performance of the Taylor expansion. In this quantity, the
second order term has a constant contribution, the fourth order term gives a linear behavior while the sixth order term adds a curvature. Note that
at small $\mu$ the magnitude of the density is small, therefore the
relative errors of ${n/( T^2 \mu) }$ are larger.

To calculate the energy density, the estimation of the LCP
and the beta function is necessary. Using zero temperature
simulations (with HMC) at $\mu=0$
with slightly shifted $\beta$ values 
on $ 24 ^3 \times 48 $ lattices the
needed mass parameters to keep the physical pion mass fixed
are found by bracketing and using a chiral perturbation theory inspired ansatz for the fitting of the pion mass dependence on the quark mass. Using finite differences we get the following results:

 \bea
     a  \left.{ \partial \beta \over \partial a} \right|_\textrm{LCP} = -0.28 \pm 0.01 ,\ 
     {\partial m\over \partial \beta } = -0.04 \pm 0.01  \textrm{ for the naive action at }\beta=5.3,\ m=0.01 \\ \nonumber
  a  \left.{ \partial \beta \over \partial a} \right|_\textrm{LCP}=-0.41\pm 0.01 ,\ 
  {\partial m\over \partial \beta } = -0.06\pm 0.01 \textrm{ for the improved action at }\beta=3.8,\ m=0.02 \\ \nonumber 
 \eea

\begin{figure}[h!]
\begin{center}
  \epsfig{file=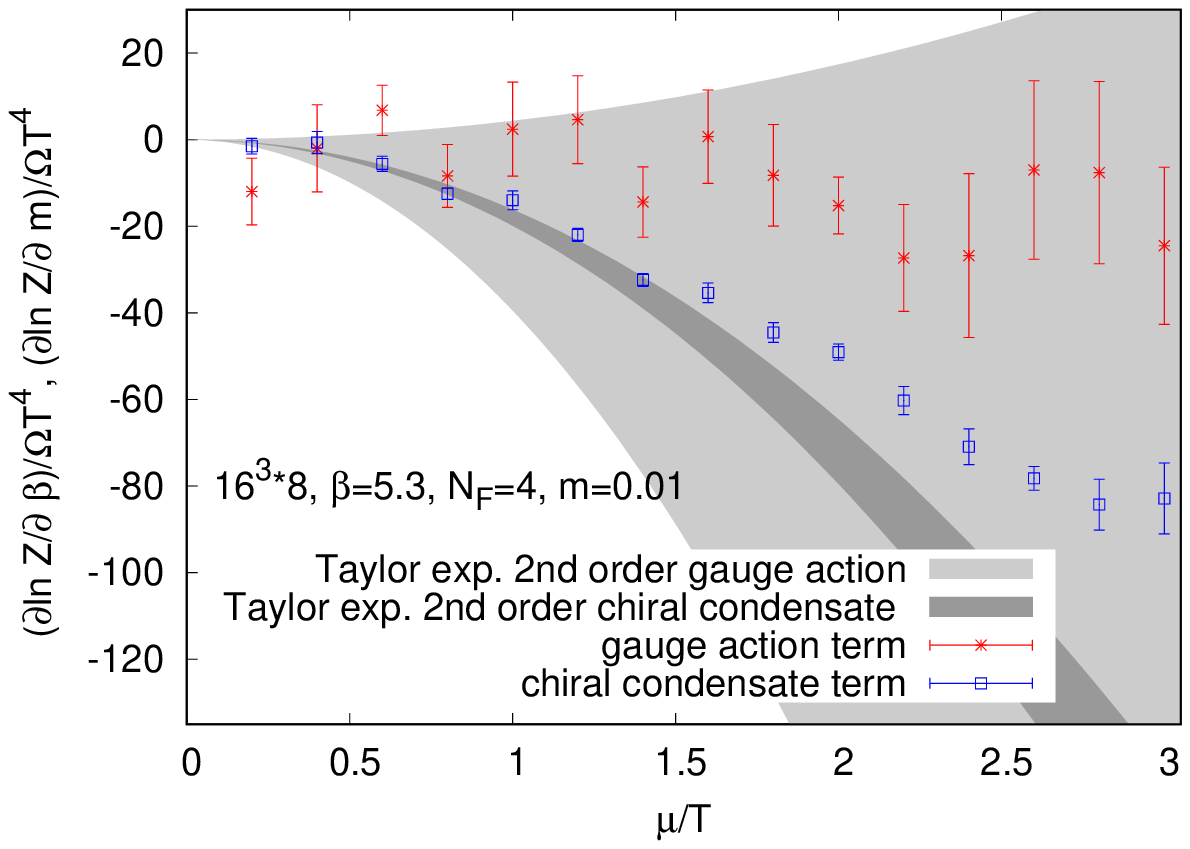, width=8.5cm}
  \epsfig{file=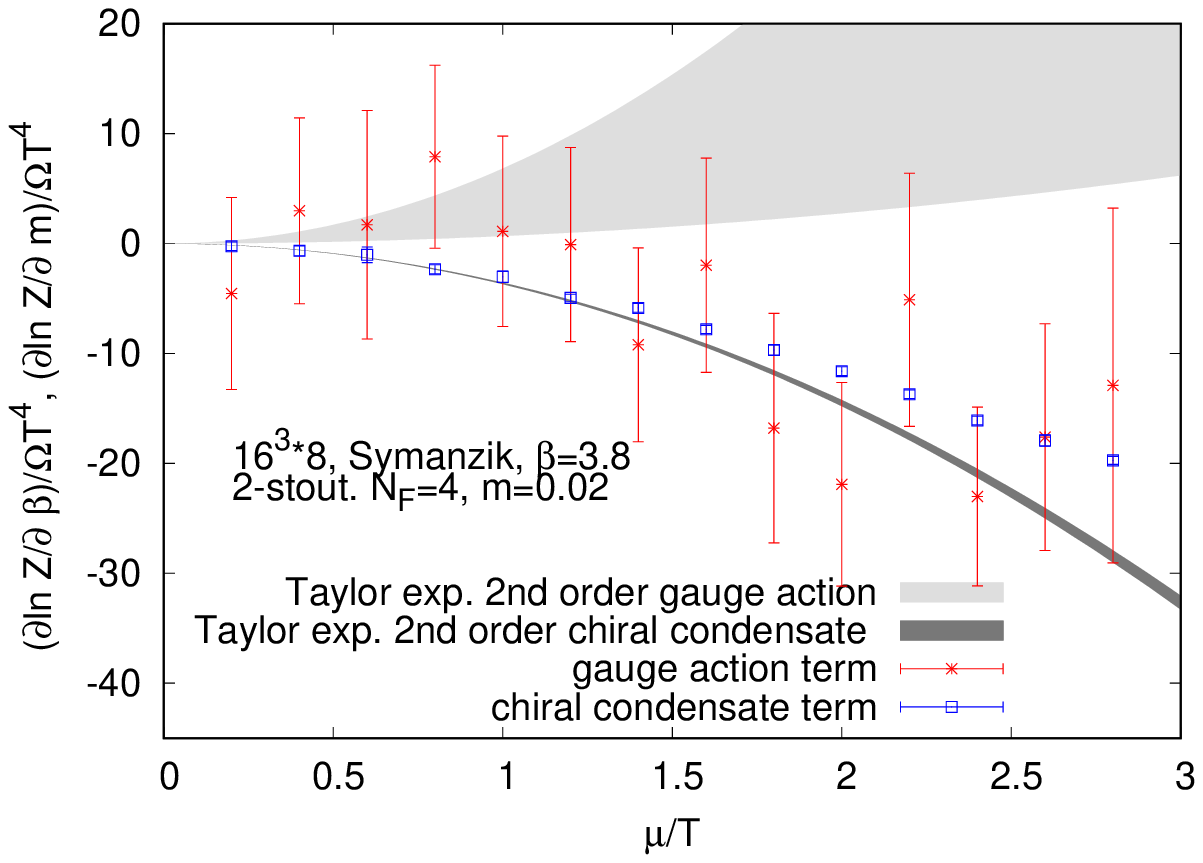, width=8.5cm}
  \caption{ The $\mu$ dependence of
    $ \partial \textrm {ln} \mathcal{Z} / \partial \beta $ and
    $ \partial \textrm {ln} \mathcal{Z} / \partial m $ 
    measured directly in CLE simulations and extrapolated
    using the 2nd order Taylor expansion from $\mu$=0.
    Using the unimproved action at $\beta=5.3$ (left) and
    Symanzik gauge action with stout smeared staggered fermions at $\beta=3.8$
    (right).
}
\label{anomaly}
\end{center}
\end{figure}

\begin{figure}[h!]
\begin{center}
  \epsfig{file=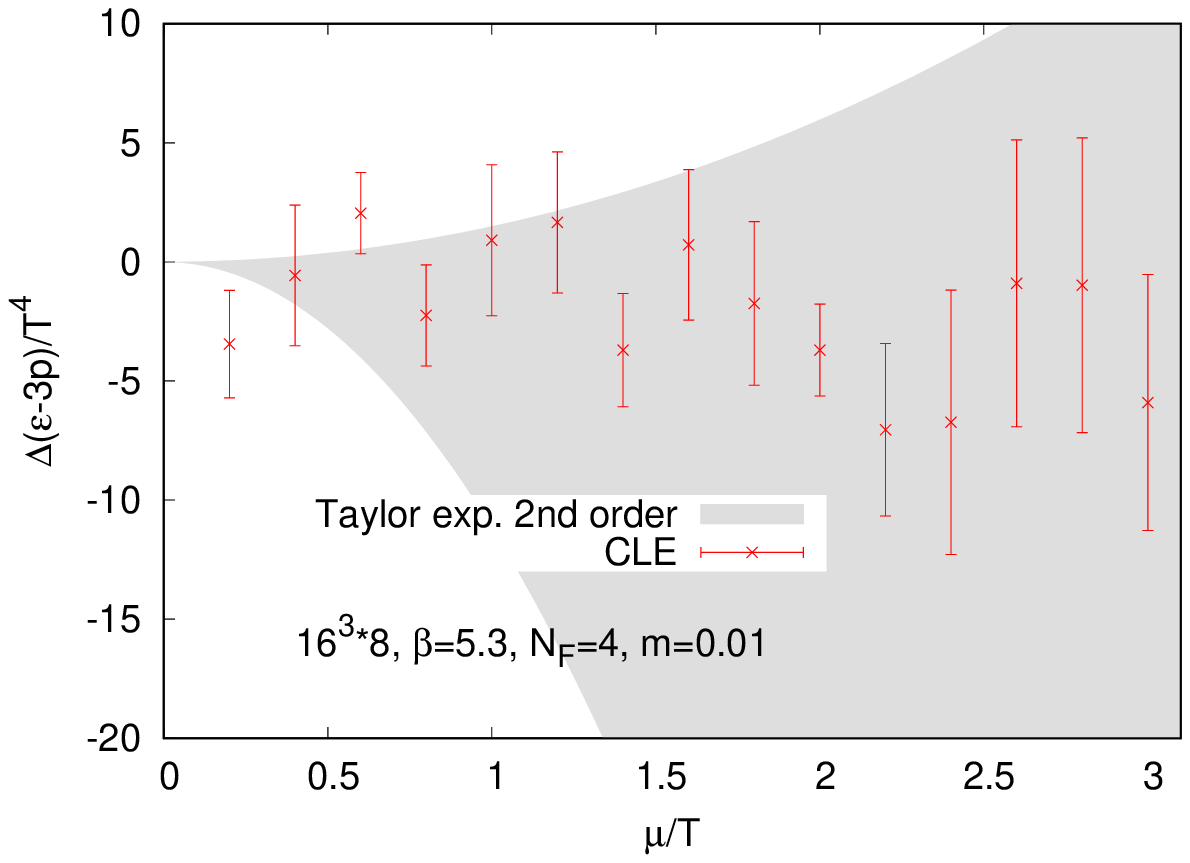, width=8.5cm}
  \epsfig{file=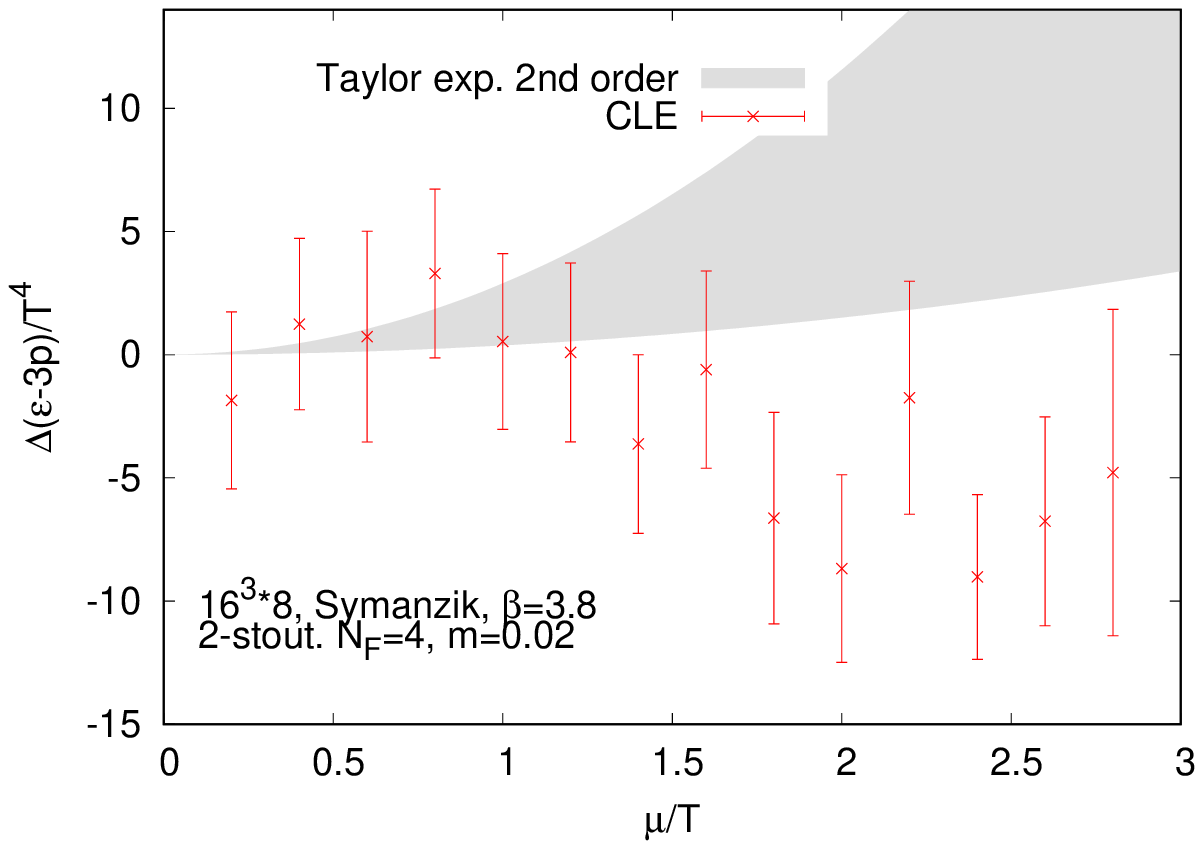, width=8.5cm}
  \caption{ The $\mu$ dependence of $ (\epsilon-3p)/T^4 $
    measured directly in CLE simulations and extrapolated
    using the 2nd order Taylor expansion from $\mu$=0.
    Using the unimproved action at $\beta=5.3$ (left) and
    Symanzik gauge action with stout smeared staggered fermions at $\beta=3.8$
    (right).
}
\label{finalanomaly}
\end{center}
\end{figure}

The $\mu$ dependence of the trace anomaly is given by a linear
combination of the $\mu$ dependence of the gauge action average
and the chiral condensate (see in eq. (\ref{traceanomaly})).
These quantities can be directly measured at nonzero $\mu$ using
CLE simulations. Alternatively their behavior can be extrapolated
using Taylor expansion from configurations at $\mu=0$.
In Fig.~\ref{anomaly} the $\mu$ dependence of
the average gauge term and the chiral condensate term from
eq. (\ref{traceanomaly}) is shown (omitting the extra factors of the beta function and the mass derivative)
for the naive action at $\beta=5.3$ as well as for the improved action
at $\beta=3.8$.  One observes that the chiral condensate term is much
better behaved with
 less fluctuations for the CLE as well as the Taylor extrapolation, and one
 observes good agreement for small chemical potential.
 The gauge action term however is much more noisy in both the CLE and
 Taylor expansion approach. With the amount of configurations at hand even the sign of the second order term is not clear. 
 In Fig.~\ref{finalanomaly} the final result for the $\mu$ dependence
 of $ (\epsilon-3p)/T^4$ including the beta function and the mass
 derivative is presented. Since the mass derivative is rather small
 the results are dominated by the gauge action term, and thus the fluctuations
 are large.  With the statistics at hand we can see that the dependence of the anomaly term on $\mu$ is significantly weaker than that of the pressure, but further conclusions are hard to gather from the data.

\section{Conclusions}
\label{concsec}

In this paper the thermodynamical properties of QCD at non-zero baryon density are studied. The aim of the study is to establish
new methods offered by the availability of the CLE simulations at
$\mu>0$ where naive importance sampling calculations are invalidated
by the sign problem. 
The results are compared to the approach relying
on the Taylor extrapolation of the results from the $\mu=0$ axis.
As the Complex Langevin equation has potential problems at smaller
temperatures related to the zeroes of the measure, here only the
high temperature phase, namely the quark-gluon plasma
state is investigated.

The pressure difference of the plasma between zero density and finite density
states is estimated using an integration method where simulations
are needed at intermediate points. The chemical potential
dependence of the trace anomaly $ \epsilon-3p$ is also calculated,
this quantity is directly accessible in a CLE simulation at $\mu>0$.

Two lattice actions were investigated, the Wilson plaquette action
with 4 flavors of naive staggered fermions and an improved action
with the Symanzik improved gauge action and stout smeared staggered fermions.
To this end the stout smearing procedure is generalized to the
complexified SL(3,$\mathcal{C}$) manifold of the link variables.
To reduce the cost of the simulations relatively heavy pion masses
 $\sim 500-700$ MeV are used.

Comparing with the usual Taylor expansion approach,
good agreement is found in the small chemical potential region
where the expansion is valid. The CLE approach
can be used to calculate at higher chemical potentials as well,
with relatively small errors.
The results suggest that the 4th order expansion formula describes the
dependence of the pressure on the chemical potential relatively closely,
while the trace anomaly $ \epsilon -3 p$ remains approximately independent
of the chemical potential for baryon chemical potentials up to $ \sim 9 T$.

The findings in this study show that the complex Langevin equation
is a useful tool to access thermodynamic quantities, and allows
calculations at high chemical potentials with small errorbars.
To allow for direct applicability for the physical world
some more work is needed:
the continuum limit and infinite volume extrapolations
still have to be carried out at the physical quark mass values.

\acknowledgments \noindent
I would like to thank Manuel Scherzer, Erhard Seiler, Ion-Olimpiu Stamatescu
for many discussions and collaboration on related topics, and
Szabolcs Bors\'anyi for discussions.
I gratefully acknowledge funding by the DFG grant Heisenberg Programme (SE 2466/1-2), 
as well as the Gauss Centre for Supercomputing (GCS) for
providing computer time on the supercomputers JURECA/BOOSTER and JUWELS
at the J\"ulich Supercomputing Centre (JSC) under the
GCS/NIC project ID HWU32.
The research was partially supported by the BMBF grant no. 05P18PXFCA.
Some parts of the numerical 
calculations were done on the GPU
cluster at the University of Wuppertal.

\bibliography{../mybib}
  
\end{document}